\documentclass[]{ica2007}
\usepackage[utf8x]{inputenc}
\usepackage[T1]{fontenc}
\usepackage{amsmath,amsfonts}
\usepackage[squaren]{SIunits}

\newcommand{\w}{\omega}
\newcommand{\derive}[1]{\frac{d {#1}}{dt}}
\newcommand{\deriven}[2]{\frac{d^{#1} #2}{dt^{#1}}}
\newcommand{\sgn}{\mathop{\mathrm{sign}}}

\begin{document}
\thispagestyle{empty}
\ICAtitle{Simulation of single reed instruments oscillations based on modal decomposition of bore and reed dynamics}
\ICApacs{43.75.Pq}
{\fontseries{m}\fontshape{n}\fontsize{10}{12}\selectfont
    Silva, F.$^1$; Debut, V.$^2$; Kergomard, J.$^1$; Vergez, C.$^1$; Deblevid, A.$^1$; Guillemain, P.$^1$}

\begin{ICAinstitutes}
\item\label{LMA} LMA UPR CNRS 7051, 31 chemin J. Aiguier, 13402 Marseille cedex 20 France; \underline{*@lma.cnrs-mrs.fr}
\item\label{ITN} Applied Dynamics Laboratory, ITN/ADL - 2686 Sacavem cedex, Portugal; \underline{vincentdebut@itn.pt}
\end{ICAinstitutes}

\ICAabstract{This paper investigates the sound production in a system made of a bore coupled with a reed valve. Extending previous work (Debut, 2004), the input impedance of the bore is projected on the modes of the air column. The acoustic pressure is therefore calculated as the sum of modal components. The airflow blown into the bore is modulated by reed motion, assuming the reed to be a single degree of freedom oscillator. Calculation of self-sustained oscillations controlled by time-varying mouth pressure and player's embouchure parameter is performed using \textsc{ode} solvers. Results emphasize the participation of the whole set of components in the mode locking process. Another important feature is the mutual influence of reed and bore resonance during growing blowing pressure transients, oscillation threshold being altered by the reed natural frequency and the reed damping. Steady-state oscillations are also investigated and compared with results given by harmonic balance method and by digital sound synthesis.}

\ICAsection{Introduction}
One of the main goals of musical acoustics is to provide models and investigation methods to improve our understanding of the behaviour of musical instruments, with application to the sound synthesis and assistance of instrument makers. In the case of woodwind instruments, 
knowledge about sound production mechanisms has improved with models including the intrinsic nonlinear coupling with the excitation. Steady-state oscillations have been investigated in frequency domain~\cite{Worman:71, Gilbert:86}, whereas time domain calculations were proposed since~\cite{Schumacher:81} for characterizing transients and stability of oscillations.
\par The aim of this paper is to present a calculation of self oscillations in time domain based on the modal decomposition of the bore input impedance. Modal analysis is widely used in musical acoustics to analyse and reproduce the vibration of complex vibrating structures, but few applications have been presented for self-sustained instruments,~\cite{Antunes:00} for the study of the bowed string, and ~\cite{Modalys} for sound synthesis. The outline is as follows : the next section presents the model of a clarinet-like system and the formulation adopted to compute self-sustained oscillations. We then present the results of calculations, and discuss the behaviour of the different components of the pressure field, and the role of reed dynamics on transients. Finally this method is compared to sound synthesis algorithm and harmonic balance technique.

\ICAsection{Model and implementation}
\ICAsubsection{Reed motion}
The elementary model, already used by many authors~\cite{Kergomard:99}, relates the volume flow entering the mouthpiece $u(t)$, and the pressure in the mouthpiece $p(t)$ to the reed displacement $y(t)$ by means of three equations characterizing the reed motion, the resonator and the flow of air through the reed channel. We assume for the characterization of the reed motion a lumped second-order mechanical oscillator with stiffness $K$ per unit area, damping $q_r$ and natural angular frequency $\w_r$, driven by the pressure drop $P_m-p(t)$ accross the reed with an inward striking behaviour :
\begin{equation}
\deriven{2}{y}+q_r\w_r \derive{y}+\w_r^2\left(y(t)-y_0\right)
=\frac{\w_r^2}{K}\left(p(t)-P_m\right),
\label{eq:reed}
\end{equation}
$P_m$ being the blowing pressure. This assumption, based on the fact that reed displacement occurs in the vertical direction mainly without torsion, has been discussed~\cite{AMvW}, and appears to be valid for non-beating reed regimes.
\ICAsubsection{The airflow}
As noted by Hirschberg~\cite{Hirschberg:99}, in the case of clarinet-like instruments, the control of the volume flow by the reed position is due to the existence of a turbulent jet. Indeed, a jet is supposed to form in the embouchure (pressure $p_\mathrm{jet}$) after the flow separation from the walls, at the end of the (very short) reed channel. Neglecting the velocity of air flow in the mouth compared to jet velocity $v_\mathrm{jet}$, the Bernoulli theorem applied between the mouth and the reed channel leads to:
\begin{equation}
P_m = p_\mathrm{jet}+\frac{1}{2}\rho v_\mathrm{jet} \quad\mbox{when $\rho$ is the air density.}
\end{equation}
Assuming a rectangular aperture of width $W$ and height $H+y(t)$ ($H$ being the height without any pressure difference), the volume flow $u$ accross the reed channel can be expressed as follows:
\begin{equation}
u(t)=W(H+y(t))\sqrt{\frac{2}{\rho}}\sqrt{P_m-p_\mathrm{jet}(t)}.
\end{equation}
Since the cross section of the embouchure is large compared to the cross section of the reed channel, it can be supposed that all the kinetic energy of the jet is dissipated through turbulence with no pressure recovery (like in the case of a free jet). Therefore, the pressure in the jet is (assuming pressure continuity) the acoustic pressure $p(t)$ imposed by the resonator response to the incoming volume flow u.

\ICAsubsection{Input impedance of the bore}
We consider a cylindrical bore (with length $L$) for the clarinet. The retained model - described extensively in~\cite{Deb04} - is the wave equation inside the tube, with Neumann and Dirichlet boundary conditions at the input and output of the bore respectively, and a source taking into account the airflow blown inside the instrument:
\begin{equation}
\begin{cases}
\left(\partial_{xx}^2 -\left(\frac{j\w}{c}+\alpha\right)^2\right)P(\w,x)
=-j\w\frac{\rho}{S}U(\w)\delta(x-x_s)
& \forall x \in \left[0,L\right] \text{ and } x_s\rightarrow 0,\\
\partial_x P(\w,x)=0 & \text{for } x=0,\\
P(\w,x)=0 & \text{for } x=L.\\
\end{cases}
\label{eq:waveequation}
\end{equation}
The pressure field $p(x,t)$ is expanded onto the modes of the air column inside the bore: $p(x,t)=\sum_{n=1}^{+\infty} f_n(x)p_n(t)$
where the family $\left\{f_n\right\}_{n\in \mathbb{N}}$ is a basis of orthogonal eigenmodes. In the case of a close/opened cylindrical pipe, $f_n(x)=\cos{(k_nx)}$, where $k_n=n\pi/2L$ and $n$ is an odd positive number. Modal coordinates $p_n(t)$ are calculated through the projection on each mode $f_n$ of equation (\ref{eq:waveequation}) in the time domain, the expansion onto the modes being truncated to the $N$ first modes:
\begin{equation}
\frac{d^2p_n}{dt^2}+2\alpha_n c \frac{dp_n}{dt}+k_n^2c^2 p_n(t)=\frac{2c}{L}Z_c\frac{du}{dt}\quad\text{for }n\in\left[1,N\right]\text{ and }Z_c=\frac{\rho c}{S}.
\end{equation}
As explained in~\cite{Deb04}, we consider that the value $\alpha$ of damping associated to each mode may be different and is frequency-independent in the neightboorhood of each angular frequency $\w_n=k_n c$. Damping is then noted $\alpha_n\simeq Y_n/L$ where $Y_n$ is the value of the admittance at $\w_n$.

\ICAsubsection{Dimensionless model}
Considering as a reference the pressure $p_M=KH$ required to close the reed channel in non-oscillating case, we introduce here the dimensionless quantities for pressure in the mouth and in the mouthpiece and for volume flow:
\begin{equation}
\tilde{p}(t)=\frac{p(t)}{p_M},\quad \tilde{p_n}(t)=\frac{p_n(t)}{p_M},
\quad \gamma=\frac{p_m(t)}{p_M},\quad \tilde{u}(t)=\frac{Z_c u(t)}{p_M}.
\end{equation}
Likewise it is convenient to define $x(t)=y(t)/H+\gamma$ and $\zeta=Z_cWH\sqrt{2(\rho p_M)^{-1}}$, such that in the case of constant control parameters $\gamma$ and $\zeta$, we have a complete model made of $N+1$ second-order \textsc{ode}:
\begin{gather}
\deriven{2}{x}+q_r\w_r \derive{x}+\w_r^2x(t)=\w_r^2 \tilde{p}(t),
\label{eq:systemedebut}\\
\frac{d^2\tilde{p_n}}{dt^2}+2\alpha_n c \frac{d\tilde{p_n}}{dt}
+(\w_n^2-\alpha^2c^2) \tilde{p_n}(t)
=\frac{2c}{L}\frac{d\tilde{u}}{dt}\quad\text{for }n\in\left[1,N\right],
\label{eq:systemecomp}\\
\text{with }\tilde{u}
=\zeta (1+x-\gamma)\sgn{(\gamma-\tilde{p})}\sqrt{|\gamma-\tilde{p}|}.
\label{eq:systemefin}
\end{gather}
It is important to enhance that parameters $\gamma$ and $\zeta$, controlled by the player while playing, are allowed to vary in time. This time dependency involves additional terms in the previous equations which allows the simulation of initial transients with a growing blowing pressure. The complete system is not written here but results are discussed in the next section.

\ICAsubsection{Beating reed}
\par Avanzini \emph{et al}~\cite{AMvW} investigated the effect of reed curling and beating against the lay of the mouthpiece, and proposed a model of variable stiffness. We include in our model the modelling of contact between lay and reed by considering an additional elastic force $F_{lay} = K_\text{lay}(H+y(t))$ when the tip gets in contact with the lay and completely closes the channel
. Further investigations would carry the case where the contact point is not the tip : as the reed curls against the mouthpiece, stiffness increases as the part of the reed which can freely move decreases, and then grows abruptly when the tip hits the lay. Another consequence of reed beating is that the volume flow vanishes when the reed closes the channel: the contact force prevents the reed from going beyond the lay in simulation, and so ensures that $1+x-\gamma\geq0$, provided $K_\text{lay}$ is high enough. This coefficient is here assumed to be an tunable parameter due to the lack of experimental investigation in this domain. Typical value is about $100$ times the stiffness of the non-beating reed.

\ICAsubsection{Numerical formulation}
The set of equations is processed using \textsc{ode} solvers. It constrains the formulation of system (\ref{eq:systemedebut}-\ref{eq:systemefin}) to be rewritten as a set of $2(N+1)$ first order \textsc{ode}, (\ref{eq:systemefin}) being used in the source term of (\ref{eq:systemecomp}). As multiple time scales are involved in the problem (duration of blowing pressure transient, bore resonance frequencies, reed resonance frequency and reed-lay contact duration), we used several \textsc{ode} solvers designed for stiff problems, in particular \texttt{ode15s} from the Matlab \textsc{ode} Suite and \texttt{lsode} from \textsc{odepack}, based on the Backward Differentiation Formula (\textsc{bfd}) methods. They lead to similar results in typical cases, for example a difference for the oscillation fundamental frequency calculated of only $2$ cents for a bore length equal to $0.5\metre$ (oscillation fundamental frequency : $f_0=170\hertz$).
\ICAsection{Results}
\ICAsubsection{Simulation of attack transients}
Using the possibility to control the mouth pressure, simulations of attack transients with increasing blowing pressure are carried out. The time evolution of $\gamma$ is as follows:
\begin{equation}
\gamma(t)=
\begin{cases}
0&\text{for }t<0,\\
\gamma_0/2 \left(1-\cos(\pi t/\tau)\right)&\text{for }t>0\text{ and }t<\tau,\\
\gamma_0&\text{for }t>\tau.
\end{cases}
\end{equation}
\begin{figure}[!ht]
    \centering
    \includegraphics[width=0.5\textwidth]{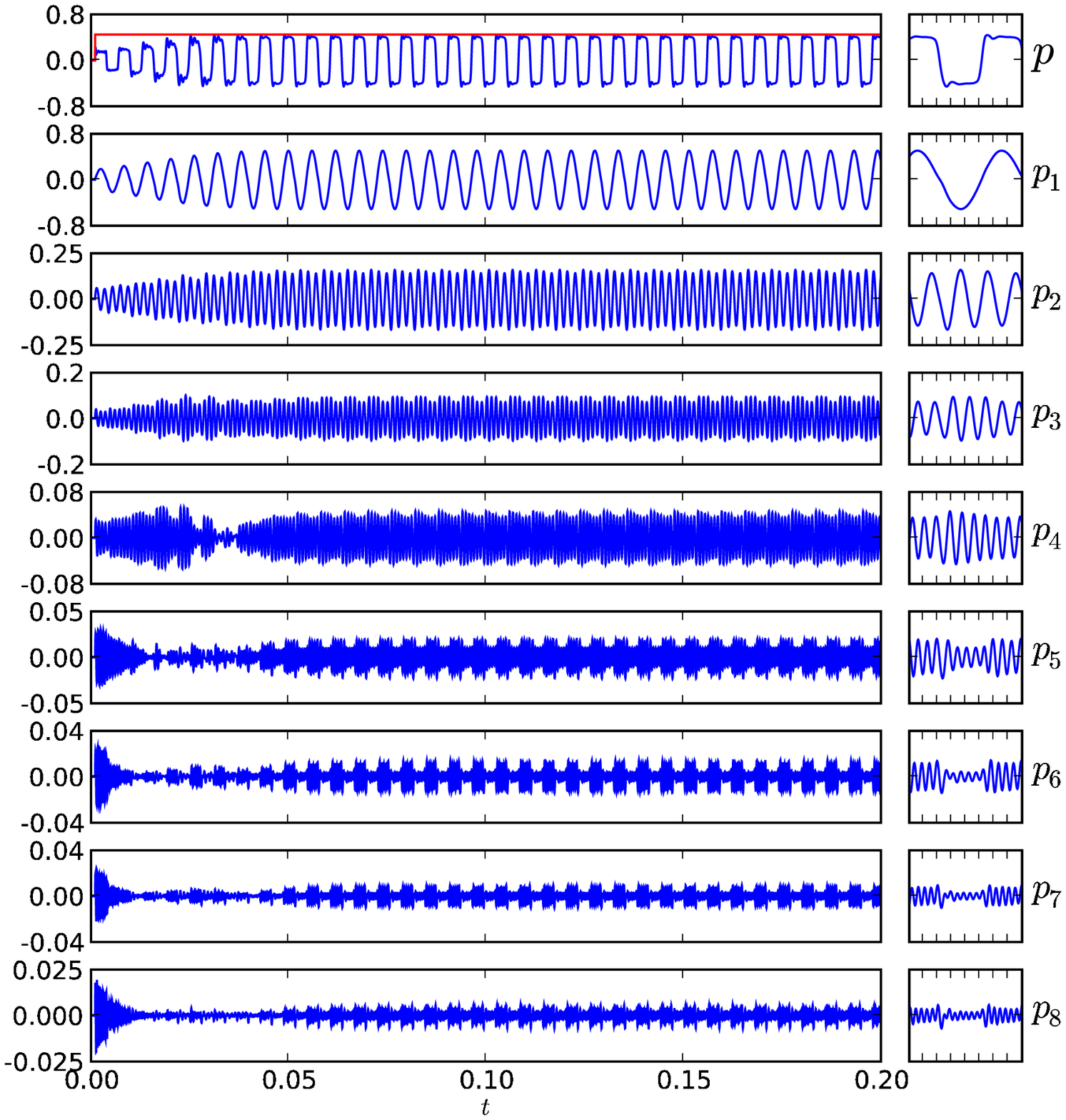}\includegraphics[width=0.5\textwidth]{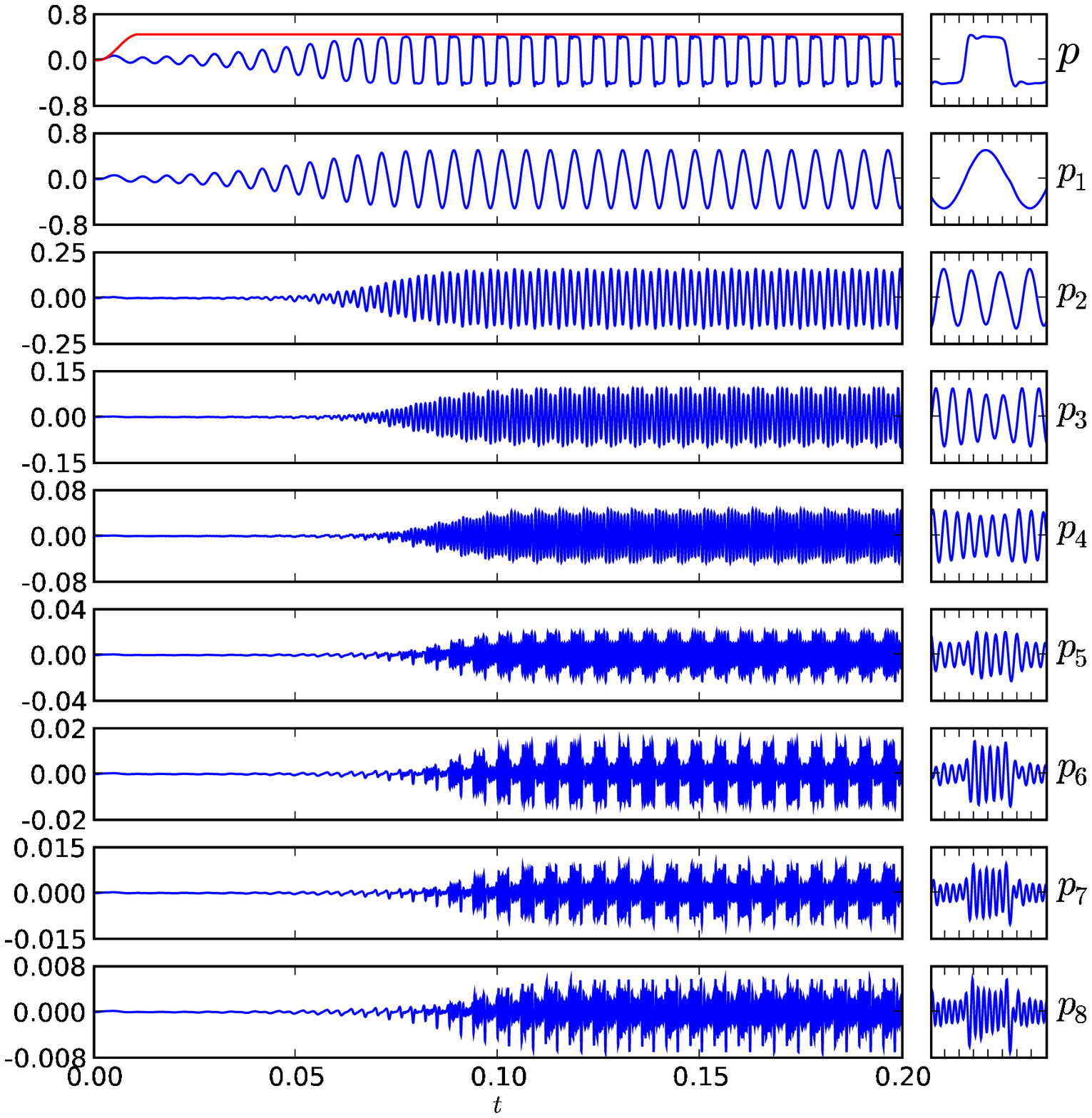}
    \caption{Pressure $p(t)$ and components $\{p_n(t)\}_{n=1,8}$ for two attack transient simulations: on the left side, instantly growing mouth pressure, and on the right, slowly growing blowing pressure ($10\milli\second$) with $qr=0.4$, $\w_r=2\pi\times 1500\hertz$ for both simulations. Waveforms in steady-state for each signal are also displayed.\label{fig:attacks}}
\end{figure}
Figure \ref{fig:attacks} shows two simulations : on the left picture, a quasi instantaneous increase of mouth pressure ($\tau=1/44100\,\second$) and in the right one a slower increase ($\tau=10\,\milli\second$). The difference between the results have to be highlighted.
On the right figure, blowing pressure slowly varying from $0$ to a value high enough to make the oscillation start: component $p_1$ is the first to appear, stabilizing its amplitude by the nonlinear phenomena in coupling. Then this saturation produces the apparition of the others components (since $t=0.05\second$). With an abruptly established mouth pressure (left figure), the whole set of components are excited, but there is a differentiation between decaying components and increasing ones. These latter are called \emph{master components}, and are involved in self-sustained oscillations generation.
On the other hand, linear stability analysis allows to conclude to the stability of decaying component for such blowing pressure, but nonlinear coupling implies energy transfer between components and the initially decaying ones finally converge to a non-zero oscillation: there are the \emph{slave components}~\cite{Deb04}.
\par Another interesting feature is the fact that higher components seems to have their amplitude modulated by fundamental oscillation (see waveforms in figure \ref{fig:attacks}), showing the difference between time domain simulation involving modal decomposition and a simple signal decomposition based on Fourier analysis: each first order resonator is excited by the input volume flow variation, that is by a rich spectrum. This results to the fact that each component is not a monochromatic signal.

\begin{figure}[!ht]
    \parbox{0.6\columnwidth}{\includegraphics[width=0.6\columnwidth]{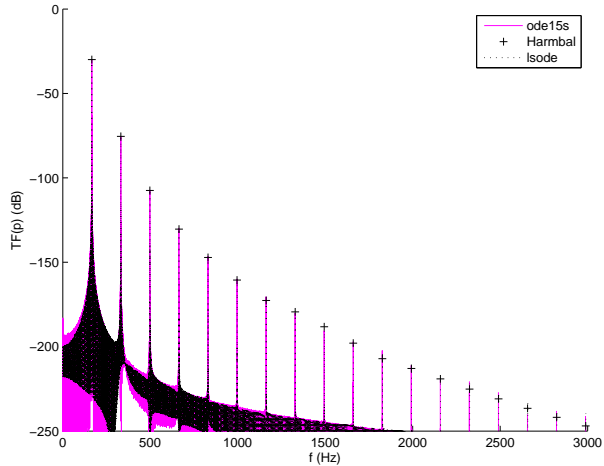}}
    \parbox{0.4\columnwidth}{\caption{Pressure $p$ power spectrum calculated with different solvers showing good agreement ($\gamma=0.45$, $\zeta=0.4$). Solvers considered include \texttt{ode15s} from the Matlab \textsc{ode} suite, \texttt{lsode} from \textsc{odepack} and \textsc{Harmbal} from \cite{Farner}. \label{f:comp_codes}}}
\end{figure}

\ICAsubsection{Comparison with Harmonic balance method}
Figure \ref{f:comp_codes} displays the result of the comparison between two time domain ode solvers ( \texttt{ode15s} from the Matlab \textsc{ode} suite and \texttt{lsode} from \textsc{odepack}) and \textsc{harmbal} (a software for computation of periodic solutions by the harmonic-balance method, \cite{Farner}) for a periodic regime. Differences between \textsc{harmbal} and time-domain solvers appear to be of the same order as differences between the two time-domain solvers. However, unlike time-domain methods, the harmonic balance method does not capture transients, therefore only periodic states can be compared.
\ICAsubsection{Comparison with sound synthesis}
The time domain scheme used for digital sound synthesis consists in obtaining digital filters representing both the reed displacement~(\ref{eq:reed}) and the input impedance. The reed displacement is discretized using centered finite difference schemes for the first and second order derivatives. This leads to a non-instantaneous response of the reed filter (i.e $y[n]$ does not depend on $p[n]$ but on $p[n-1]$). The impedance is modeled by a digital filter expressing, at sample $n$, the pressure $p[n]$ as: $p[n]=b_{co}u[n]+V[n]$, where $V[n]$ includes past (and known) values of $p$ and $u$. For the purpose of this paper,  this digital filter is computed from the inverse Fourier transform of the reflexion coefficient associated to the modal impedance. These discretizations allow to solve explicitely the nonlinear coupling, whatever the bore geometry, as it has been shown in \cite{Guillemain:05}. Since the modal impedance comes from a truncated serie of modes and does not take into account radiation losses or tonehole network, it is worth noting that at high frequencies, the associated reflection coefficient tends to $1$ and the admittance tends to infinity. This behavior generates instabilities in the acoustic flow during attack transients and we are working on this problem.
\begin{figure}[!ht]
    \parbox{0.6\columnwidth}
    {\includegraphics[width=0.6\columnwidth]{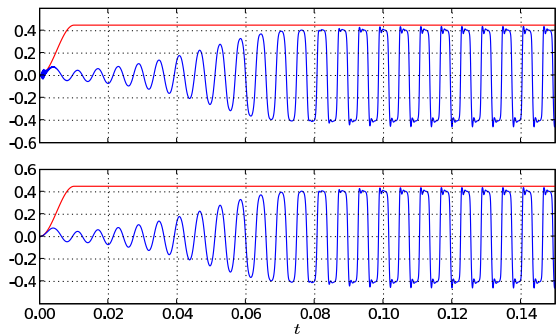}}
    \parbox{0.4\columnwidth}{\caption{Pressure $p$ signal resulting from sound synthesis (top picture) and simulation (bottom picture) for a increasing blowing pressure ($\gamma_0=0.45$ and $\tau=10\milli\second$).}}
\end{figure}
\ICAsection{Conclusion}
The approach presented in this paper relies on a time domain formulation, leading to ab initio calculations. Therefore, different stable regimes are computable depending obviously on initial conditions but also on the possibly time-varying values of control parameters. Moreover the modal decomposition of the dynamics allows an explicit writing of the problem using meaningful parameters in musical acoustics such as natural resonance frequencies and dampings. Since these quantities are accessible to measurements, our simulations can be reasonably easily compared to experiments. This is not always the case with time domain methods using impulse responses or reflection functions for example~\cite{Schumacher:81}.
\par Another advantage of the time-domain modal approach is that the resulting formulation (a system of first-order \textsc{ode}s) is  the same as the one used in dynamical system studies. Benefits are obvious to analyze behaviors observed in simulations (for example the analysis of attack transients in terms of master/slave components in section \textbf{Results} of this paper). A more practical consequence is that many reliable solvers are available for first-order \textsc{ode} systems. This allows to test many routines without additional development, in order to check the validity of a simulation result, or in order to look for the best solver for a particular problem. For example, this appeared to be decisive when considering reed beating regimes, very demanding numerically, which required a routine specifically designed for stiff problems.
\par The computational cost may appear high compared to other implementations. This is not a limitation of the time domain modal formulation since other similar approaches are used for real-time sound synthesis~\cite{Modalys}, but is mainly due to the type of chosen solver (variable order, multi-step scheme). Our approach is thus definitely not intended for sound synthesis. On the other hand, our simulations are expected to be more precise and aimed at investigating model dynamics.
\ICAsubsection{Acknowledgments}
The study presented in this paper was lead with the support of the
French National Research Agency \textsc{anr} within the
\textsc{Consonnes} project.

\bibliographystyle{ICAbib1}
\bibliography{BibComplete}
\end{document}